\documentclass{sf2a-conf2023}
\usepackage{graphicx}
\usepackage{hyperref}
\usepackage[]{natbib}  
\usepackage{epstopdf}
\usepackage[dvipsnames]{xcolor}
\usepackage[utf8]{inputenc}

\def\BibTeX{{\rm B\kern-.05em{\sc i\kern-.025em b}\kern-.08em
    T\kern-.1667em\lower.7ex\hbox{E}\kern-.125emX}}
\bibpunct{(}{)}{;}{a}{}{,}  

\begin{document}

\TitreGlobal{SF2A 2023}

\title{Environmental transition: overview of actions to reduce the environmental footprint of astronomy}

\runningtitle{Environmental transition}

\author{L. Leboulleux}\address{Univ. Grenoble Alpes, CNRS, IPAG, 38000 Grenoble, France}
\author{F. Cantalloube}\address{Aix Marseille Univ, CNRS, CNES, LAM, Marseille, France}
\author{M.-A. Foujols}\address{Institute Pierre-Simon Laplace (IPSL), Sorbonne University, CNRS, Paris, France,}
\author{M. Giard}\address{Institut de Recherche en Astrophysique et Plan\'etologie, Universit\'e de Toulouse, CNRS, CNES, Toulouse, France}
\author{J. Guilet}\address{Universit\'{e} Paris-Saclay, Universit\'{e} Paris Cité, CEA, CNRS, AIM, 91191, Gif-sur-Yvette, France}
\author{J.~Knödlseder$^4$}
\author{A.~Santerne$^2$}
\author{L.~Todorov$^2$}
\author{D.~Barret$^5$}
\author{O.~Berne$^4$}
\author{A.~Crida}\address{Univ. C\^ote d'Azur, Observatoire de la C\^ote d'Azur, CNRS, Lagrange (UMR7293), 06304 Nice, France}
\author{P.~Hennebelle$^5$}
\author{Q.~Kral}\address{LESIA, Observatoire de Paris, Université PSL, CNRS, Sorbonne Université, Univ.~Paris Diderot, Meudon, France}
\author{E.~Lagadec$^6$}
\author{F.~Malbet$^1$}
\author{J.~Milli$^1$}
\author{M.~N'Diaye$^6$}
\author{F.~Roques$^{7}$}



\setcounter{page}{237}


\maketitle

\begin{abstract}
To keep current global warming below $1.5^\circ$C compared with the pre-industrial era, measures must be taken as quickly as possible in all spheres of society. Astronomy must also make its contribution. In this proceeding, and during the workshop to which it refers, different levers of actions are discussed through various examples: individual efforts, laboratory-level actions, impact evaluation and mitigation in major projects, institutional level, and involvement through collectives.
\end{abstract}




\section{Introduction}
The Paris Climate Agreement, ratified by all countries during the COP21 in 2015, aims to keep global warming at $1.5^\circ$C-$2.0^\circ$C compared to the pre-industrial era. This objective is based on strategies to reduce greenhouse gas emissions, mainly due to the use of fossil fuels, which must be implemented at all levels of society. The French community of astronomy has also taken up this issue, by multiplying greenhouse gas emissions assessments and actions to limit its own impact on global warming: a dozen astronomy laboratories have calculated their greenhouse gas emissions. Several initiatives have been launched, as a result of these diagnostics, including: (i) founding collectives of academic researchers, such as \textit{Labos 1point5}\footnote{\href{https://labos1point5.org}{https://labos1point5.org}}, a cross-disciplinary group in France \citep{Benari2023}, and \textit{Astronomers for Planet Earth}\footnote{\href{https://astronomersforplanet.earth/}{https://astronomersforplanet.earth/}}, an international movement of astronomy students, educators, amateurs and scientists, \citep{Cantalloube2021, Wagner2023, Stevens2023}, (ii) setting up hybrid or remote meetings and conferences \citep{Kral2021, Sarabipour2021, Moss2021}, (iii) introducing a system for tracking greenhouse gas emissions when booking business trips (at LESIA), etc.

In 2023, during the annual conference of the Soci\'et\'e Fran\c{c}aise d'Astronomie et d'Astrophysique (SF2A), we organized the third edition of a special session dedicated to the environmental transition within the professional astronomy community. In this special half-day session, we addressed the question of how our research projects are evolving in the context of the Paris Climate Agreement. The aim of the discussions was to address the opportunities and implications for individuals, laboratories, institutions and major scientific projects, as well as the impact and recognition of these actions.

The content of this proceeding is broken down according to the topics addressed by the speakers, each focusing on a different level of action: institutional level with Martin Giard (Deputy director Head of Astronomy \& Astrophysics Division of the CNRS, section~\ref{s:MG}), involvement through collectives with Marie-Alice Foujols (CNRS engineer at IPSL in climate science numerical modeling, core member of Labos1point5, section~\ref{s:MAF}), assessing and mitigating the impact of major astronomical projects with Jürgen Knödlseder (researcher at IRAP, author of \emph{The carbon footprint of astronomical research infrastructures}, section~\ref{s:JK}), laboratory-level actions with Lilia Todorov (ITA CNRS in the administrative department, leader of the sustainability group at LAM, section~\ref{s:LT}) and Alexandre Santerne (researcher at LAM, member of Labos1point5), 
and individual implication in collective actions with Jérôme Guilet (researcher at CEA, member of Scientist Rebellion, section~\ref{s:JG}).

\section{Institutional level: the french astronomy and astrophysics community prospective}
\label{s:MG}

In 2024, as every five years, the Centre National de Recherche Scientifique (CNRS) will realize a new community prospective exercise in the field of astronomy and astrophysics. This will be the occasion to collectively decide the french scientific priorities in the domain and our roadmap for the period 2025-2030. The situation of our professional activities in the society will also be reviewed in detailed with a  focus on three different topics attached to social progress within our community and in relation with our social and ecological environment: 1) Equity, Diversity, and Inclusion, 2) Carbon and ecology transition, and 3) astronomy and territories and participative science. Even regarding the yet ongoing work at CNRS and in laboratories, this prospective and its preparatory discussions and team work plan to evaluate, support, encourage and, guide the ecological transition of research. Between others, the impact of research and formation as a tool for social progress will be discussed, in addition with the gestion  of decarbonation of already existing projects and infrastructures and the ongoing development of new ones.

\section{Collectives: the success of Labos1point5}
\label{s:MAF}

Labos1point5 \citep{Benari2023} is a pioneering project started in 2019 in France by a collective of academic researchers, whose goal is to better understand and reduce the environmental impact of research, especially on the Earth’s climate. One of the main highlight of Labos 1point5 is the development of an online open source tool to assess the carbon footprint of a research institute \emph{GES 1point5}\footnote{\href{https://labos1point5.org/ges-1point5}{https://labos1point5.org/ges-1point5}} \citep{Mariette2022}. The associated tool  \emph{Scenario 1point5} enables mitigation scenarios to be explored by means of more than 15 reduction measures in the various areas. The networking platform \emph{Transition 1point5} enables labs in transition to exchange ideas. To date, over 800 research institutes in France have used \emph{GES 1point5}, representing thousands of employees, producing a unique database that can be studied to assess in detail the origin and determinants of the carbon footprint of French research. Several papers related to the carbon footprint of research using the \emph{GES 1point5} database and the Labos 1point5 survey have been published or submitted for publication in international peer reviewed journals \citep{Mariette2022, berne2022carbon, de2023purchases}. 
\emph{GES 1point5} is currently being adapted in the USA with expected applications at the Space Telescope Science Institute and NoirLab. 

Since 2022, Labos 1point5 is an official research consortium supported by several national institutions in France: CNRS, ADEME, INRAE and INRIA. Labos 1point5 is also a think tank, organizing monthly webinars and producing texts and reports on various topics relating to the ecological transition in research and teaching. Among other actions, Labos 1point5 sent to national recruitment committees in France a text in which it is asked that environmental activities are taken into consideration in the hiring and career evaluation.

\section{Major astronomical projects: example of CTAO}
\label{s:JK}

The Cherenkov Telescope Array Observatory (CTAO) is the next generation ground-based instrument for gamma-ray astronomy at very high energies. With 64 telescopes located in the northern and southern hemispheres, the CTAO will be the first open ground-based gamma-ray observatory and the world’s largest and most sensitive instrument to study high-energy phenomena in the Universe.

Recently, an Office for the Reduction of the Environmental Footprint of CTAO was created with the mission of ensuring that the environmental footprint of CTAO does not exceed the planetary limits within which humanity can continue to develop and thrive for generations to come \citep{Richardson2023}. The Office was established under the initiative of the CTAO Managing Director and approved by the administrative Council of the Observatory. Bringing the Office into life is however challenging, since the environmental footprint is conceived as an additional constraint on top of a wealth of managerial and technical challenges that are already tricky to cope with in a complex international environment. Also, it is difficult to obtain the financial resources needed to commission studies or recruit staff in a context where project costs are already tightly constrained.

A first rough estimate of sources of greenhouse gas emissions reveals that CTAO operations are likely to dominate the life-cycle carbon footprint, due to the observatory’s 30 years lifetime. Electricity consumption will be a major source of emissions during operations, underscoring the need to implement renewable energy solutions. Telescope construction is likely to be another major source of emissions, which is challenging to control owing to their provision in form of in-kind contributions.

To go beyond this first rough estimate, it is planned to perform an environmental life cycle assessment (LCA) of the observatory, in order to develop plans for powering CTAO using renewable energy sources, as well as to conduct optimisation efforts that will reduce the environmental footprint of CTAO during the construction and operations phases. Hopefully, the financial resources will soon become available to start these activities.

\section{Laboratory-level actions: focus on LAM}
\label{s:LT}

The Laboratoire d’Astrophysique de Marseille is a joint research unit (UMR 7326) under the supervision of the CNRS, the Aix-Marseille University and the CNES. The sustainable development group was created in Novem- ber 2018, with about 10 volunteer staff from different backgrounds: doctoral students, researchers, lecturers, engineers and administrative staff. The aim of this group approach is to implement solutions to effectively re- duce the impact of our activities on the environment and make daily life in the laboratory more eco-responsible. 

The above-mentioned actions are structured around several themes: 1) From 2020 onwards, the group has set itself the target of quantifying its greenhouse gas emissions using the methodology common to all laboratories proposed by Labos1point5. The group intends to continue this analysis over the next few years in order to implement a multi-year action plan aimed at reducing the LAM’s carbon footprint. The results were presented at a LAM seminar and an article is currently being prepared (Santerne et al., in prep.). 2) The group writes 2-3 articles in the laboratory’s monthly science newsletter, highlighting the actions and informing the staff. To raise awareness and popularise the issue, the group organises seminars at the laboratory with external speakers. It also provides an in-house training for staff through \emph{Ma Terre en 180 minutes} activity \citep{gratiot2023transition}. 3) To promote biodiversity and engage against soil artificialization in the laboratory, a communal vegetable garden was created in the LAM’s patio. 4) The group has introduced recycling bins for paper, cardboard, ink cartridges and IT equipment. To combat the misuse of plastic, the group has also worked to eliminate plastic cups and bottles, as well as installing water fountains and distributing water bottles, mugs and coffee cups.

\section{Civil disobedience}
\label{s:JG}

Non-violent civil disobedience is widely considered to be an efficient means of achieving social and political change. It has therefore been used for more than a century by citizens fighting for various causes (antislavery, women or black people rights, anticolonialism, ecology, etc.). The failure of other means of actions to foster a reduction of greenhouse gas emission has pushed a growing number of citizens and scientists to turn to non- violent civil disobedience in the fight against climate change. The active engagement of scientists in such actions can be justified by several reasons. Recent developments in the philosophy and sociology of science show that neutrality is neither attainable nor desirable \citep{Fragnière2022}. Scientists need to engage actively with society because producing and supplying information is not sufficient to ensure that this information will be taken into account and produce adequate changes \citep{Oreskes2022}. The failure of climate mitigation is caused in large parts by opposing political and economical interests \citep{Stoddard2022} that make power relationships unavoidable in the fight against climate change. Finally, protests against fossil fuel projects have been shown to be more effective when they include multiple strategies including civil disobedience \citep{Thiri2022}. More and more academic publications therefore call for scientists to engage in civil disobedience \citep{Capstick2022} and for the universities, research institutions and unions to support researchers that do so \citep{Gardner2021}. Scientist Rebellion, and its french branch \textit{Scientifiques en r\'ebellion} \footnote{\href{https://scientifiquesenrebellion.fr}{https://scientifiquesenrebellion.fr}}, is an international collective of scientists funded in 2020 that engages in non-violent civil disobedience in order to demand emergency decarbonisation and degrowth, facilitated by wealth redistribution.

\section{Discussions and conclusions}

The multi-scale panel covered by J. Guilet, L. Todorov, J. Knödlseder, M. Giard, and M.-A. Foujols during the SF2A special session S21 prompted many questions and discussions with the audience. Among others, the topic of decision-making and decision-makers was discussed, with the example of the carbon footprint of business trips: the audience and speakers pointed out a profound discrepancy between 1) what is expected or required from researchers (e. g. PhD candidates and postdoctoral researchers feel compelled to travel a lot to attend conferences and give talks in order to increase their visibility and foster collaborations with a view to obtaining grants, scholarships or permanent positions), 2) the resources provided to reduce the trip’s carbon footprint (e. g. the booking platform does not offer train trips abroad and/or trains travel is more expensive than flights...), 3) individual and collective determination to actively reduce our carbon footprint.

In this context, which decisions could be made and how to implement them, with what level of compromise? The first option is bottom-up, driven by employees, as is the case in many laboratories. But sometimes, opposition to change can also stem from other employees: for instance at LERMA, carbon quota could not be implemented since the adoption of this measure required the support of a large majority of employees ($>80\%$ of employees). 
Conversely, other actions can be imposed by employers or research infrastructures on the community, for instance, before using data from the Hawaiian telescope, watching informational videos about sacred land use in Hawaiian observatories despite protests from the local population. For reference, the Labos 1point5 collective has compiled a list of actions carried out in laboratories, together with the processes for obtaining their approval.

The weight of research infrastructures was mentioned, along with the idea that today’s science has become highly specialized, dependent on research institutes that produce large quantities of greenhouse gases and are only accessible to an exclusive few. Others spoke of the need for citizens to regain access to knowledge through participatory science. 

At a less structural level, the discussions evoked the weight of individual decisions and actions: for instance, protesting or civil disobedience involves only the citizen, not their employer. Some junior researchers attending the round table expressed their frustration and concern that a career in astronomy does not seem compatible with a reduction of the individual or collective carbon footprints. These young researchers also reported suffering from solastalgia (also known as eco-anxiety).

In the end, there seems to be a general desire for change at various levels in our community. It was indeed one of the most attended SF2A special sessions, with an explicitly large proportion of junior researchers, despite parallel sessions about future large astronomical facilities that may have attracted more audience. To conclude, one can quote J. Knödlseder, explaining that this huge challenge ahead of us is also an opportunity for change: "\textit{when did we have such an opportunity to start from a blank page and write a new chapter in history?}". As a community, we sincerely hope that decisions will be rapidly taken and implemented in the years to come, at all possible levels, in order to do our part to achieve the goal set by the Paris Climate Agreement and limit the impact of astronomical research on the health of the planet.

\begin{acknowledgements}
The authors thank the SF2A organization committees (SOC and LOC) who made this conference a large success. The SOC of the Environmental Transition session is also grateful for the panel of speakers for their implication during the talks and during the discussions with the audience, and to Pierre Kern for taking and sharing notes.
\end{acknowledgements}

\bibliographystyle{aa}
\bibliography{leboulleux} 

\begin{thebibliography}{18}
\expandafter\ifx\csname natexlab\endcsname\relax\def\natexlab#1{#1}\fi

\bibitem[{{Ben Ari}(2023)}]{Benari2023}
{Ben Ari}, T. 2023, Nature Reviews Physics

\bibitem[{Bern{\'e} {et~al.}(2022)Bern{\'e}, Agier, Hardy, Lellouch, Aumont,
  Mariette, \& Ben-Ari}]{berne2022carbon}
Bern{\'e}, O., Agier, L., Hardy, A., {et~al.} 2022, Environmental Research
  Letters, 17, 124008

\bibitem[{{Cantalloube} \& {Burtscher}(2021)}]{Cantalloube2021}
{Cantalloube}, F. \& {Burtscher}, L. 2021, in SF2A-2021: Proceedings of the
  Annual meeting of the French Society of Astronomy and Astrophysics. Eds.: A.
  Siebert, 310--312

\bibitem[{Capstick {et~al.}(2022)Capstick, Thierry, Cox, Berglund, Westlake, \&
  Steinberger}]{Capstick2022}
Capstick, S., Thierry, A., Cox, E., {et~al.} 2022, Nat. Clim. Chang. 12,
  773–774

\bibitem[{De~Paepe {et~al.}(2023)De~Paepe, Jeanneau, Mariette, Aumont, \&
  Estevez-Torres}]{de2023purchases}
De~Paepe, M., Jeanneau, L., Mariette, J., Aumont, O., \& Estevez-Torres, A.
  2023, bioRxiv, 2023

\bibitem[{Fragnière(2022)}]{Fragnière2022}
Fragnière, A. 2022, Université de Lausanne

\bibitem[{Gardner {et~al.}(2021)Gardner, Thierry, Rowlandson, \&
  Steinberger}]{Gardner2021}
Gardner, C.~J., Thierry, A., Rowlandson, W., \& Steinberger, J.~K. 2021,
  Frontiers in Sustainability, 2

\bibitem[{Gratiot {et~al.}(2023)Gratiot, Klein, Challet, Dangles, Janicot,
  Candelas, Sarret, Panthou, Hingray, Champollion,
  {et~al.}}]{gratiot2023transition}
Gratiot, N., Klein, J., Challet, M., {et~al.} 2023, PLOS Sustainability and
  Transformation, 2, e0000049

\bibitem[{{Kral}(2021)}]{Kral2021}
{Kral}, Q. 2021, in SF2A-2021: Proceedings of the Annual meeting of the French
  Society of Astronomy and Astrophysics. Eds.: A. Siebert, 317--319

\bibitem[{{Mariette} {et~al.}(2022){Mariette}, {Blanchard}, {Bern{\'e}},
  {Aumont}, {Carrey}, {Ligozat}, {Lellouch}, {Roche}, {Guennebaud},
  {Thanwerdas}, {Bardou}, {Salin}, {Maigne}, {Servan}, \&
  {Ben-Ari}}]{Mariette2022}
{Mariette}, J., {Blanchard}, O., {Bern{\'e}}, O., {et~al.} 2022, Environmental
  Research: Infrastructure and Sustainability, 2, 035008

\bibitem[{Moss {et~al.}(2021)Moss, Adcock, Hotan, Kobayashi, Rees, Si{\'e}gel,
  Tremblay, \& Trenham}]{Moss2021}
Moss, V.~A., Adcock, M., Hotan, A.~W., {et~al.} 2021, Nature Astronomy, 5, 213

\bibitem[{Oreskes(2022)}]{Oreskes2022}
Oreskes, N. 2022, Proc. Indian Natl. Sci. Acad., 88(4):824–8

\bibitem[{Richardson {et~al.}(2023)Richardson, Steffen, Lucht, Bendtsen,
  Cornell, Donges, Drüke, Fetzer, Bala, von Bloh, Feulner, Fiedler, Gerten,
  Gleeson, Hofmann, Huiskamp, Kummu, Mohan, Nogués-Bravo, Petri, Porkka,
  Rahmstorf, Schaphoff, Thonicke, Tobian, Virkki, Wang-Erlandsson, Weber, \&
  Rockström}]{Richardson2023}
Richardson, K., Steffen, W., Lucht, W., {et~al.} 2023, Science Advances, 9,
  eadh2458

\bibitem[{Sarabipour {et~al.}(2021)Sarabipour, Khan, Seah, Mwakilili, Mumoki,
  S{\'a}ez, Schwessinger, Debat, \& Mestrovic}]{Sarabipour2021}
Sarabipour, S., Khan, A., Seah, Y. F.~S., {et~al.} 2021, Nature Human
  Behaviour, 5, 296

\bibitem[{{Stevens} \& {Moss}(2023)}]{Stevens2023}
{Stevens}, A.~R.~H. \& {Moss}, V.~A. 2023, Communicating Astronomy with the
  Public Journal, 32, 15

\bibitem[{Stoddard {et~al.}(2021)Stoddard, Anderson, Capstick, Carton,
  Depledge, Facer, Gough, Hache, Hoolohan, Hultman, H\"{a}llstr\"{o}m, Kartha,
  Klinsky, Kuchler, L\"{o}vbrand, Nasiritousi, Newell, Peters, Sokona,
  Stirling, Stilwell, Spash, \& Williams}]{Stoddard2022}
Stoddard, I., Anderson, K., Capstick, S., {et~al.} 2021, Annual Review of
  Environment and Resources, 46, 653

\bibitem[{Thiri {et~al.}(2022)Thiri, Villamayor-Toms, Scheidel, \&
  Demaria}]{Thiri2022}
Thiri, M.~A., Villamayor-Toms, S., Scheidel, A., \& Demaria, F. 2022,
  Ecological Economics, 195, 107356

\bibitem[{{Wagner} {et~al.}(2023){Wagner}, {Mingo}, {Majidi}, {Gokus},
  {Burtscher}, {Kayhan}, {Kobayashi}, {Mehta}, {Moss}, {Ossenkopf-Okada},
  {Rice}, {Stevens}, {Waratkar}, \& {Woods}}]{Wagner2023}
{Wagner}, S.~M., {Mingo}, B., {Majidi}, F.~Z., {et~al.} 2023, Nature Astronomy,
  7, 244

\end{thebibliography}

\end{document}